# Note: Quasi-real-time analysis of dynamic near field scattering data using a graphics processing unit

G. Cerchiari,[1,2] F. Croccolo,[1,a)] F. Cardinaux,[1] and F. Scheffold[1,2]

[1]*Department of Physics, University of Fribourg, Ch. Du Musée 3, 1700 Fribourg, Switzerland*
[2]*Fribourg Center for Nanomaterials, University of Fribourg, 1700 Fribourg, Switzerland*

We present an implementation of the analysis of dynamic near field scattering (NFS) data using a graphics processing unit (GPU). We introduce an optimized data management scheme thereby limiting the number of operations required. Overall, we reduce the processing time from hours to minutes, for typical experimental conditions. Previously the limiting step in such experiments, the processing time is now comparable to the data acquisition time. Our approach is applicable to various dynamic NFS methods, including shadowgraph, Schlieren and differential dynamic microscopy.

Light scattering techniques are powerful tools to characterize the structure and dynamics of simple and complex fluids.[1] Structural information about the system is contained in the angular dependence of the scattered intensity, while its temporal fluctuations are characteristic of the dynamics. Traditional static and dynamic light scattering techniques analyze scattering in the far field, but the same information can also be obtained under near field conditions.[2,3] Near field scattering (NFS) techniques are nowadays available in different layouts using radiation sources ranging from coherent[4] and partially coherent light,[5-7] to white light[8] or X-rays[9] with different experimental constraints. The ease of access to extremely low scattering angles or the possibility to perform spatially resolved experiments are interesting features of NFS that have allowed studies on liquids,[2-5] colloidal dispersions[10] or biological samples.[11] A drawback of NFS is the heavy computational load involved in the evaluation of data.[4,8,12] Typically, a sequence of thousands of mega-pixel images is necessary to evaluate both the static and dynamic spectra with sufficient accuracy.

In this study, we present an implementation of dynamic NFS using a graphics processing unit (GPU). The highly parallel architecture of GPUs is optimized to execute a certain operation (*kernel*) on multiple data elements (*stream*) simultaneously. It results in an acceleration of processing originally used to handle computer graphics. The use of stream processing for non-graphical applications is expanding due to the availability of high-level programming interfaces[13,14] and has been exploited in applications ranging from computational biology, cryptography, computational chemistry, biomedical imaging or optics.[13-17] Very recently Lu and coworkers have reported on the application of an algorithm similar to ours for the rapid analysis of fluorescence microscopy images of liquid colloidal suspensions.[17] Here we show that for dynamic NFS a remarkable speed-up can be achieved when processing the data using a GPU instead of a CPU. Moreover we introduce an effective method to manage the graphics random access memory (G-RAM) in order to optimize the number of operations and data transfers in the treatment of NFS data. Overall, we are able to reduce the processing time from hours to minutes, for our typical experimental settings. Previously the limiting step in such experiments, the time required for data processing is now comparable to the actual data acquisition time.

As a representative test experiment, we present NFS results for concentration non-equilibrium fluctuations[18] in a binary fluid subjected to a thermal stress.[4,5,19] The sample is a mixture of tetrahydronaphthalene and n-dodecane ($c$=50% w/w) contained in a sapphire flat cell with a gap thickness of 1.3mm. Non-equilibrium fluctuations are driven by a temperature difference of 20°C ($T_{avg}$=25°C) across the sample. The structure and dynamics of the concentration fluctuations are characterized using a heterodyne NFS instrument having a shadowgraph detection layout.[5,6] A low coherence light source emitting at 680±10nm (Super Lumen Diodes, Broad Lighter S680) is expanded to a diameter of 20mm and collimated. The cell is centered in the optical axis with its surface normal to the axis. Finally, a charge-coupled device (Vosskühler, CCD4000) placed at a distance of 26cm away from the cell acquires intensity-images at rate of 4Hz. Raw images are composed of a superposition of static contributions (dominated by the primary beam) and a time dependent contribution arising from the scattering due to concentration fluctuations. A sample image is presented in Fig.1(a) wherein the beam profile as well as residual interferences due to reflections and stray light are visible. In a scattering experiment, the structure and the dynamics of the system under study are characterized by the wave vector dependence of the scattered intensity $I_s(q)$ and the

---

[a)] Electronic mail: fabrizio.croccolo@unifr.ch

intermediate scattering function $f(q,\tau)$ (ISF), respectively.[4,5] An elegant way to obtain both quantities from NFS experiments is to use to the double-frame differential analysis.[4,5] In a first step, images are normalized by their spatial average values and differences $\Delta i_m(x,\tau) = i_m(x,t) - i_m(x,t+\tau)$ between normalized images separated by a time delay $\tau$ are computed. The resulting spatial distributions of intensity show a homogeneous speckle pattern.[18] This is illustrated in Fig.1 for a time delay of (b) $\tau = 0.25s$, (c) and 25s. The scattered intensity and the ISFs are obtained by analyzing the time delay dependence of power spectra of $\Delta i_m(x,\tau)$:

$$\langle |\Delta I_m(q,\tau)|^2 \rangle = 2a\{T(q)I_s(q)[1-f(q,\tau)] + B(q)\}, \quad (1)$$

where $\Delta I_m(q,\tau)$ is the spatial Fourier transform of $\Delta i_m(x,\tau)$, $a$ is a renormalization constant, $T(q)$ is the transfer function of the imaging optics,[5,6] and $B(q)$ is the noise background.[4,5] A 2-D power spectrum for $\tau = 25s$ is presented in Fig.1(d). For the usual case of stationary dynamics, ensemble averaging $\langle ... \rangle$ in (1) can be performed over the measurement time. A selection of azimuthally averaged power spectra is presented in Fig.1(e) and (f).

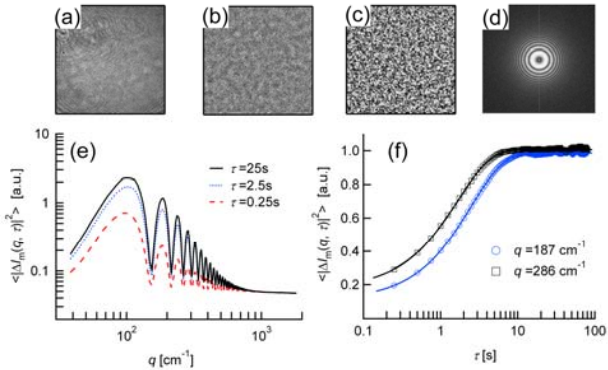

FIG. 1. (Color online). *Upper panels:* (a) Shadowgraph image, image differences with time delay of (b) 0.25s and (c) 25s, (d) spatial power spectrum of (c). *Lower panels:* Power spectra as a function of (e) wave vector q and (f) time delay τ.

The processing of NFS data using the double-frame differential analysis of Eq. 1 requires performing a limited number of consecutive operations on large data sets. As we will show a remarkable increase in processing speed can be achieved using a parallel processing approach. An equally important aspect of a fast processing is to avoid redundant calculations. For a number $N$ of images, the direct calculation of the power spectra for all the available $N-1$ values of $\tau$ involves $N(N-1)/2$ iterations. However, some redundant calculations can be avoided. In particular, one can take advantage of the linearity of the Fourier transform by evaluating FFT's on single images instead of image differences, thus reducing their number to $N$. A further speedup can be obtained by dividing the spatial Fourier transform of each image by its zero-frequency value for image normalization. Finally, only the square moduli of FFT differences are evaluated $N(N-1)/2$ times. These processing steps require images to be read from the motherboard random access memory (RAM), transfer it to the G-RAM for the GPU-processing, FFT-transform, calculate differences and finally calculate and store the average for any given $\Delta N$ in the memory. For the case of CPU-processing the images are read directly from the RAM. However for the case of GPU computing data has to be transferred from the RAM to the limited G-RAM. The amount of available G-RAM therefore sets a ceiling up to which GPU-processing provides full computational benefits. As a matter of fact, if the number of images to be analyzed exceeds the G-RAM capacity then data has to be transferred multiple times leading to a performance decrease. In the latter case data transfer can be optimized by organizing the memory as a First-In-First-Out (FIFO) queue. In order to obtain a 'general-purpose' code able to treat any amount of images of any size, no *a priori* optimization has been built into the code. Here we chose to optimize memory allocation prior to starting the calculations. For a memory capacity for only $N^* < (N-1)$ images we can write $N-1 = \alpha N^* + \beta$, were $\alpha, \beta$ are positive integer numbers and $\beta < N^*$. The program needs to calculate all the possible FFT differences step by step. To this end $N^*$ images are uploaded into the FIFO, one image is uploaded into the 'current image area' (see Fig.2), all possible differences are calculated and the results are averaged and stored in the 'differences area'. These steps are repeated until the first triangle of matrix elements has been processed. Next one diagonal is filled by subsequently replacing one FIFO image and uploading one 'current image'. This procedure is repeated for all the $\alpha$ diagonals and subsequently the values for the remaining upper-right triangle are computed. For simplicity we discuss only the case $\beta = 0$ and we find that the number of image transfer operations to the G-RAM is $N_{FFT} = N \cdot (2\alpha - 3 + 2/\alpha) \rightarrow 2\alpha N$ for $\alpha \gg 1$ and thus much smaller compared to the total number of differences computed $\cong N^2/2$. Our reference motherboard is equipped with an 8-core CPU (INTEL, XEON X3440 at 2.53GHz) and a GPU-board (NVIDIA, Tesla C2050). The double-frame differential analysis is implemented both in C++/CUDA for the GPU-based code and in C++ for the CPU-based one. The performances of the two versions are compared in order to quantify the speed increase due to the GPU. The total execution time of the program is the key parameter for evaluating its performance. However, we also analyzed separate timings for different operations.[20] With both codes we processed different sets of images with different sizes, spanning from 32 up to 2048 images of $32 \times 32$ up to $4k \times 4k$ pixels.

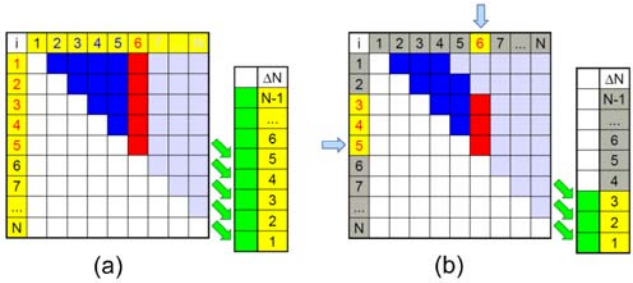

FIG. 2. Organization schemes for the memory in the two distinct cases (a) when all the data can be stored in the G-RAM, and (b) when the data exceeds the G-RAM capacity. The vertical axis denotes image FFT's, the horizontal one denotes the image FFT to compute differences for a given $\Delta N$, and the green column contains the calculated power spectra for corresponding time delays $\tau = \Delta N \times 0.25s$. Blue elements of the matrix stand for already calculated differences, while red elements are calculated at the current iteration. Cyan arrows indicate images that are uploaded at the current iteration

We now focus our attention to the ratio between the GPU and the CPU processing time $t_{CPU}/t_{GPU}$. The result of this analysis is shown in Fig.3 as a contour plot of this ratio (z-axis) as a function of image size (x-axis) and number of images (y-axis). For essentially all of our experimental conditions the GPU code provides a decrease of the processing time, the only exceptions being a combination of images size smaller than $128\times128$ and $N<128$. The highest gain in speed reaches $t_{CPU}/t_{GPU} = 32$ for $512\times512$ sized images and $N = 2048$ images. Under these conditions, while the CPU code requires about 47 minutes for analyzing data, the same analysis is performed by the GPU one in less than 2 minutes. As a general trend we note that increasing the number of images provides an increased speed-up ratio up to image sizes of $512\times512$. This is certainly related to the fact that for 2048 images of $512\times512$ pixels the amount of data slightly exceed the G-RAM capacity, therefore for 2048 images of larger size multiple loading is preventing larger speedups. A more detailed analysis of the timing of the different operations is provided in the supplementary material.[20]

From this we conclude that the performance of the code is limited by the competition between the speed increase due to the GPU parallelization and the slow down due to the additional time needed to transfer data to the GPU. This finding can provide guidelines for the further optimization of the code for specific applications. For example, in our case we chose to analyze all the time delays and to store the entire 2D power spectra, which might be unnecessary in many cases. Relaxing one or more of these requirements would of course improve the performance for large data sets. A widely used concept would be, for example, to consider only delay times distributed evenly on a logarithmic scale.[21]

We are grateful to S. Biffi for help with MatLab code, to R. Caravita for his cThread C++ class, to R. Cerbino and L. Chantada for setting up an earlier variant of the NFS experiment and to R. A. Armenta-Calderón for interesting discussions. G.C., on leave from the University of Milan, would like to thank the Fribourg Center for Nanomaterials for financial support during a summer internship. F.C. acknowledges EU financial support: Marie Curie Contract IEF-251131, DyNeFI. Financial support by the Swiss National Science Foundation (Project Nr. 132736) and the Adolphe Merkle Foundation is gratefully acknowledged.

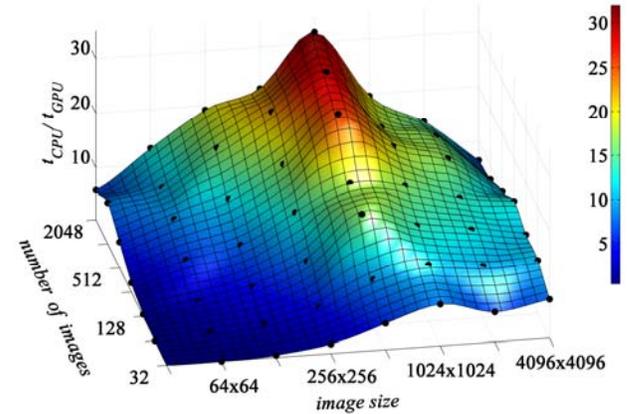

FIG. 3. (Color online). Ratio between the total processing time of the GPU and the CPU based code as a function of the size of the images and their number.

[1]B. J. Berne and R. Pecora, *Dynamic Light Scattering* (Wiley, New York, 1976).
[2]M. Giglio, M. Carpineti, and A. Vailati, Phys. Rev. Lett. **85**, 1416 (2000).
[3]R. Cerbino and A. Vailati, Curr. Opin. in Colloid Interface Sci. **14**, 416 (2009).
[4]F. Croccolo, D. Brogioli, A. Vailati, M. Giglio, and D. S. Cannell, Appl. Opt. **45**, 2166 (2006).
[5]F. Croccolo, D. Brogioli, A. Vailati, M. Giglio, and D. S. Cannell, Phys. Rev. E **76**, 41112 (2007).
[6]F. Croccolo and D. Brogioli, Appl. Opt. **50**, 3419 (2011).
[7]D. Brogioli, D. Salerno, F. Croccolo, R. Ziano, and F. Mantegazza, New J. Phys. **13**, 123007 (2011).
[8]R. Cerbino and V. Trappe, Phys. Rev. Lett. **100**, 188102 (2008).
[9]R. Cerbino, L. Peverini, M. A. C. Potenza, A. Robert, P. Bösecke, and M. Giglio, Nature Physics **4**, 238 (2008).
[10]A. Oprisan, S. Oprisan, and A. Teklu, Appl. Opt. **49**, 86 (2010).
[11]L. G. Wilson, V. A. Martinez, J. Schwarz-Linek, J. Tailleur, G. Bryant, P. N. Pusey, and W. C. K. Poon, Phys. Rev. Lett. **106**, 018101 (2011).
[12]D. Magatti, M. D. Alaimo, M. A. C. Potenza, and F. Ferri, Appl. Phys. Lett. **92**, 241101 (2008).
[13]M. Garland, S. L. Grand, J. Nickolls, J. Anderson, J. Hardwick, S. Morton, E. Philips, Y. Zhang, and V. Volkov, Micro **28**, 13 (2008).
[14]S. Che, M. Boyer, J. Meng, D. Tarjan, J. W. Sheaffer, and K. Skadron, J. Parallel Distrib. Comput. **68**, 1370 (2008).
[15]P. J. Lu, H. Oki, C. A. Frey, G. E. Chamitoff, L. Chiao, E. M. Fincke, C. M. Foale, S. H. Magnus, W. S. M. Jr., D. M. Tani, P. A. Whitson, J. N. Williams, W. V. Meyer, R. J. Sicker, B. J. Au, M. Christiansen, A. B. Schofield, and D. A. Weitz, J. Real-Time Image Proc. **5**, 179 (2010).
[16]T. Shimobaba, Y. Sato, J. Miura, M. Takenouchi, and T. Ito, Opt. Express **16**, 11776 (2008).
[17]P. J. Lu, F. Giavazzi, T. E. Angelini, E. Zaccarelli, F. Jargstorff, A. B. Schofield, J. N. Wilking, M. B. Romanowsky, D. A. Weitz, and R. Cerbino, Phys. Rev. Lett. **108**, 218103 (2012).
[18]A. Vailati and M. Giglio, Nature **390**, 262 (1997).
[19]F. Croccolo, H. Bataller, and F. Scheffold, (unpublished).
[26]See supplementary material at [url will be inserted by aip] for a more detailed analysis of timings for different operations".
[21]K. Schätzel, M. Drewel, and S. Stimac, J. Modern Optics **35**, 711 (1988).